# Atomic-scale mechanisms for magnetostriction in $CoFe_2O_4$ and $La_{0.5}Sr_{0.5}CoO_3$ oxides determined by differential x-ray absorption spectroscopy

G. Subías[1,*], V. Cuartero[2], J. García[1], J. Blasco[1], S. Pascarelli[3]

[1]*Instituto de Ciencia de Materiales de Aragón, Departamento de Física de la Materia Condensada, CSIC-Universidad de Zaragoza. C/ Pedro Cerbuna 12, E-50009 Zaragoza, Spain*

[2] *Centro Universitario de la Defensa, Carretera de Huesca s/n, 50090 Zaragoza, Spain*

[3] *ESRF-The European Synchrotron, 71 Avenue des Martyrs, Grenoble (France)*

[*]Email: gloria@unizar.es

**Abstract**

The atomic environments involved in the magnetostriction effect in $CoFe_2O_4$ and $La_{0.5}Sr_{0.5}CoO_3$ polycrystalline samples have been identified by differential extended x-ray fine structure (DiffEXAFS) spectroscopy. We demonstrate that cobalt atoms at octahedral sites are the responsible for their magnetostriction. The analysis of DiffEXAFS data indicates that the local-site magnetostrictive strains of Co atoms are reversed in these two oxides in agreement with the macroscopic magnetostriction. For the $CoFe_2O_4$ spinel, a large negative strain along the (100) direction have been determined for the $CoO_6$ octahedron causing a tetragonal contraction in contrast with the $La_{0.5}Sr_{0.5}CoO_3$ perovskite, where a positive moderate strain along the (100) direction was found resulting in a tetragonal expansion. The different local-site magnetostriction is understood in terms of the different valence and spin state of the Co atoms for the two oxides. The macroscopic magnetostriction would be explained then by the relative change in volume, either contraction in $CoFe_2O_4$ or expansion in $La_{0.5}Sr_{0.5}CoO_3$, when the tetragonal axis of the Co site is reoriented under an externally applied magnetic field.

# I. Introduction

Magnetostriction (MS) is a phenomenon observed in all ferromagnetic materials, which develop a mechanical deformation when they become magnetized by the application of an external magnetic field. It is a consequence of the magnetoelastic coupling and it pertains to the strain produced along the field direction. Functional materials with giant MS play an important role in a broad range of industrial, biomedical and defense applications since they can reversibly convert energy between the magnetic and elastic states [1]. Within these materials, alloys based on Fe, Ni and Co transition metals in combination with certain rare-earth elements like Tb and Dy have been intensively studied due to their giant room-temperature magnetic field induced strain up to thousands of ppm [2]. In these rare-earth-based materials, the magnetostrictive strain arises from the coupling of the orbital anisotropic 4f charge distribution with both the 4f magnetic moment (via the spin-orbit coupling) and the local distortions (via the crystal field) as it has been recently determined using differential x-ray absorption spectroscopy (DiffXAS) in the $TbFe_2$ alloy [3].

In order to develop alternative magnetostrictive materials with reduced cost, easy manufacturing and enhanced properties, oxide based magnetic materials like ferromagnetic spinels and perovskite-type oxides have been also studied for a long time although the magnitude of MS strain is not as high as in the rare-earth compounds. In the oxide-based materials, strong electron-lattice coupling is considered as responsible for the giant MS but different mechanisms can apply depending on the class of magnetic oxides. Among the perovskite-type oxides, both mixed-valence magnetoresistance manganites [4] and $La_{1-x}Sr_xCoO_3$ cobaltites [5] show giant magnetostrictive effects. In the former, the observed volume MS at Tc is related to the quenching of the charge localization under the application of an external magnetic field whereas in the latter, the observed anisotropic MS is thought to arise from the orbital instability of $Co^{3+}$ ions under the applied magnetic field inducing a transition from orbital non-degenerate low-spin state to an orbitally degenerated intermediate-spin state. However, these large strains are observed only at very low temperatures and require high magnetic fields, which restrict the use of these materials for practical applications for sensing and actuation. On the other hand, spinel structured ceramic oxides such as cobalt ferrites offer a wide range of interesting properties for application due to its large anisotropic MS of hundreds of ppm in both single-crystal and polycrystalline samples at room temperature and low magnetic fields [6, 7].

Cobalt ferrites ($Co_xFe_{3-x}O_4$) can be considered as a cobalt-substituted variant of magnetite ($FeFe_2O_4$). It has been recently investigated for important technological applications both in its stoichiometric ($CoFe_2O_4$) and cation substituted forms [8, 9]. It has a partially inverse spinel structure where the cobalt ions occupy the B-sites (octahedral local symmetry) of the crystal lattice. The changes in the magnetic properties of magnetite due to cobalt substitution for iron, $Co_xFe_{3-x}O_4$, are due to the differences in the properties of $Co^{2+}$ and $Fe^{2+}$ (for $x \leq 1$). The high MS of cobalt ferrites has been associated to the properties of the $Co^{2+}$ cation and it was explained within the single-ion crystal-field model [6, 7, 10-12], which considers that Co cations occupy the octahedral B-sites in the spinel lattice, the orbital angular momentum of $Co^{2+}$ transition metal is unquenched and couples to both the spin momentum and the tetragonal distortion of the octahedral crystal field. Therefore, MS increases with the cobalt content as it was experimentally demonstrated [7].

Despite of the extensive study of these ferrites, there is not any direct experimental measurement of the local distortion of any of the two atom sites originated by field-induced magnetization because this distortion is expected to be too small to be detected by conventional experimental techniques such as x-ray or neutron diffraction. As it is well known, the best technique able to determine the local structure around a specific atom in a solid is x-ray absorption spectroscopy (XAS) [13, 14]. However, its sensitivity to determine local distortions is far from the expected one to be able to measure the strain coming from MS though it can be improved by the subtraction of XAS spectra. In fact, DiffXAS [15, 16] has demonstrated sensitivity to the atomic displacements of the order of femtometers and it has made the direct assessment of local-scale MS possible, as reported for several metal alloys [17, 18].

We present here a DiffXAS study of the atomic-scale MS of the $La_{0.5}Sr_{0.5}CoO_3$ perovskite and $CoFe_2O_4$ spinel ferrite. $La_{0.5}Sr_{0.5}CoO_3$ is a ferromagnet below 250 K [19, 20] with a high anisotropic MS as large as 1000 ppm below 100 K but at relatively high magnetic field of 14 T [5]. At moderate fields of 2 T, the magnetostrictive strain has been reported to be ~ 450 ppm. On the other hand, $CoFe_2O_4$ is a ferrimagnet with a high Curie temperature $T_C$= 793 K and the value of the anisotropic magnetostrictive strain varies from 200 to 400 ppm at room temperature and low magnetic fields (**B**<1T) for polycrystalline samples [7, 21, 22]. We expect to extract the magnitude and sign of the changes in individual bond lengths in order to determine which atomic strains are responsible for the macroscopic MS effects in the two samples.

## II. Experimental Details

Polycrystalline powder samples of $La_{0.5}Sr_{0.5}CoO_3$ and $CoFe_2O_4$ have been synthesized by a sol-gel method using the citrate route as described elsewhere [5, 23]. The obtained samples were single phase as confirmed by conventional x-ray powder diffraction. Magnetic measurements agreed with samples with the right oxygen stoichiometry and reveal that $CoFe_2O_4$ is saturated at room temperature for a magnetic field of 2 T [23, 24] and magnetic saturation is also achieves at this field for $La_{0.5}Sr_{0.5}CoO_3$ at low temperatures well below Tc ≈ 250 K [5, 25].

XAS and DiffXAS measurements were performed at beamline BM23 of the European Synchrotron Radiation Facility [26]. Pellets were prepared by dilution with cellulose to get a jump of about 1 at the absorption edge. Different pellets will be optimized for $CoFe_2O_4$ at the Co and Fe K-edges, respectively. For the DiffXAS experiment, pellets were inserted into a He flow cryostat that was placed at the center of a 2 T magnetic field wheel, which rotated in the plane perpendicular to the propagation vector of the x-rays. This allows us to perform measurements from 10 K up to room temperature. The absorption coefficient, $\mu(E)$, was recorded with the magnetic field direction **B** parallel and then perpendicular, $\mu^{\parallel}(E)$ and $\mu^{\perp}(E)$, respectively, to the x-ray electric-field polarization vector **E**. The differential absorption, $\Delta\mu(E) = \mu^{\parallel}(E) - \mu^{\perp}(E)$, averaged over a large number of cycles to improve the signal-to-noise ratio, was normalized by dividing by the edge jump. Spectra were measured in transmission in the quick-scanning mode of the monochromator. The acquisition time for each XAS spectrum was of about 40 s with each pair of measurements repeated 400 times and averaged to minimize statistical noise. Error in $\Delta\mu(E)$ is estimated from DiffXAS measurements made at +45° and -45° angles between the magnetic field and the x-ray electric-field polarization vector. To analyze the DiffXAS signal, we first perform a standard Extended X-ray Absorption Fine Structure (EXAFS) data analysis to obtain the average local structure of Co and Fe atoms, respectively. The EXAFS analysis was performed by means of Athena and Artemis programs of the Demeter package [27]. All the spectra were fitted in the k range available for the DiffXAS spectra, from 2 to 10 Å$^{-1}$ for

$CoFe_2O_4$ and from 2.5 to 9.5 Å$^{-1}$ for $La_{0.5}Sr_{0.5}CoO_3$ with free parameters the interatomic distance (R) and the Debye-Waller factor ($\sigma^2$) for each coordination shell. Constraints in coordination numbers are imposed by the particular crystallographic structures.

### III. Results

#### A. $CoFe_2O_4$

Figure 1 shows the normalized DiffXAS spectra $\Delta\mu(E)$ at both the Co K and the Fe K edges at room temperature for the $CoFe_2O_4$ sample. The experiment was repeated but the data acquisition was started with a 90º phase shift with respect to the first measurements. The DiffXAS signals are inverted, as shown in Fig.1, as expected for a magnetostrictive signal. The normalized x-ray absorption spectra are also shown multiplied by 0.007 (Co K-edge) and 0.001 (Fe K-edge) for comparison purposes. Additional control measurements were performed with the magnetic field oriented at ± 45º with respect to the x-ray electric-field polarization vector yielding no structure and further demonstrating that the DiffXAS signals are originated by local-strain MS. As we can see, the room temperature DiffXAS signal at the Co K edge is about eight times larger than that at the Fe K edge indicating that the maximum local distortion occurs at the Co site. At both absorption edges, a large dichroic signal is also observed at the absorption onset, of peak-peak amplitude $|\Delta\mu| \sim 2.5\times10^{-2}$ at the Co K edge and $|\Delta\mu| \sim 2\times10^{-3}$ at the Fe K edge. In contrast to more itinerant metallic ferromagnets [28], these large dichroic signals have been already observed at the Co and Fe $L_{2,3}$ edges in $CoFe_2O_4$ due to the localized nature of the magnetic moments.

Figure 2 shows Co K edge and Fe K edge Differential Extended X-ray Absorption Fine Structure (DiffEXAFS) signals at room temperature and at 40 K. The overall shape of the DiffEXAFS signals is similar at both temperatures. However, the amplitude of the DiffEXAFS oscillations at the Fe K edge slightly increases at low temperatures whereas it is independent of temperature at the Co K edge. At 40 K, the Co K-edge DiffEXAFS signal is only about three times larger than that at the Fe Kedge.

#### B. $La_{0.5}Sr_{0.5}CoO_3$

Figure 3a show the normalized DiffXAS spectra $\Delta\mu(E)$ at the Co K at T=25 K. The experiment was repeated but the data acquisition was started with a 90º phase shift with respect to the first measurements. The DiffXAS signals are inverted, as shown in Fig.3, as expected for a magnetostrictive signal. The normalized x-ray absorption spectra are also shown multiplied by 0.0015 for comparison purpose. A dichroic signal is also observed at the Co K absorption onset, $|\Delta\mu| \sim 1.8\times10^{-3}$ but it is smaller than for the cobalt ferrite and opposite in sign. Figure 3b compares Co K edge DiffEXAFS signals at 25 K and 75 K. The amplitude of the DiffEXAFS oscillations at 75 K slightly decreases (by about 20%) in agreement with the temperature dependence of the macroscopic anisotropic MS that is nearly constant below 100 K [5].

## IV. Theory and Analysis

The expression for the polarized EXAFS fine-structure function in anisotropic systems such as single-crystals, oriented powders or layered compounds is given by

$$\chi(k) = \sum_j A_j(k) \cdot \sin(k \cdot s_j + \varphi_j(k)) \cdot e^{-k^2 \sigma_j^2} \cdot 3 \cos \theta_j^2 \qquad (1)$$

where $A_j(k)$ and $\varphi_j(k)$ are the scattering amplitude and phase functions of the photoelectron, $s_j$ is the scattering path length that is equivalent to $2R_j$ (distance between the absorbing and neighbour atoms) for single scattering, $\sigma_j^2$ is the Debye-Waller factor and $\theta_j$ is the angle between the x-ray polarization vector $\vec{E}$ and the $R_j$ vectors. The DiffEXAFS spectrum obtained by the difference between two EXAFS spectra measured applying a magnetic field parallel and perpendicular to the x-ray polarization vector, respectively, is small so it is possible to express the DiffEXAFS signal in terms of a first order Taylor expansion of the x-ray fine-structure function (Eq. 1), with respect to the modulated parameter.

$$\Delta\chi(k) = 3 \cdot \sum_j A_j(k) \cdot e^{-k^2 \sigma_j^2} \cdot \left[ k \cdot \cos\left(k \cdot s_j + \varphi_j(k)\right) \cdot \Delta s_j \cdot \cos \theta_j^2 - 2 \cdot k^2 \cdot \sin\left(k \cdot s_j + \varphi_j(k)\right) \cdot \Delta \sigma_j^2 \cdot \cos \theta_j^2 - \sin 2\theta_j \cdot \Delta \theta_j \cdot \sin\left(k \cdot s_j + \varphi_j(k)\right) \right] \qquad (2)$$

Therefore, $\Delta\chi(k)$ contains three contributions. The first corresponds to the strain acting on each interatomic distance $\Delta s_j$ and shows a cosine phase dependency so it is in quadrature with the EXAFS signal. The second term retains the sine phase dependency of the original EXAFS signal and describes changes to the disorder $\Delta \sigma_j^2$. In our particular case where we have performed the measurements at fixed temperature and always under an applied magnetic field, we can consider that the contribution from the disorder term is negligible. Moreover, we have also considered that there is not change on the bond angle $\Delta \theta_j$ induced by the magnetic field so the third term is discarded too. Then, we analyse the DiffEXAFS data by considering exclusively the strain term so Eq. (2) results in

$$\Delta\chi(k) = \sum_j A_j(k) \cdot e^{-k^2 \sigma_j^2} \cdot k \cdot \cos\left(k \cdot s_j + \varphi_j(k)\right) \cdot \Delta s_j \cdot 3 \cos \theta_j^2 \qquad (3)$$

We note here that (1) For each individual $j^{th}$ coordination shell, the related DiffEXAFS signal is in quadrature with its respective EXAFS signal and (2) $\Delta s_j$ is two times the difference in the interatomic distance between the absorbing atom and an atom in the $j^{th}$ coordination shell for the two orientations of the applied magnetic field, parallel and perpendicular to the x-ray polarization vector so it depends on the angle between the scattering path $s_j$ and the magnetic field. For a polycrystalline sample, the DiffEXAFS spectrum will be

$$\Delta \chi(k) = \sum_j A_j(k) \cdot e^{-k^2 \sigma_j^2} \cdot k \cdot \cos\left(k \cdot s_j + \varphi_j(k)\right) \cdot \Delta s_j' \qquad (4)$$

where $\Delta s_j' = <3 \cdot \cos \theta_j^2 \cdot \Delta s_j> = \frac{1}{4\pi} \int_{\theta=0}^{\pi} \int_{\phi=0}^{2\pi} 3 \cdot \cos \theta_j^2 \cdot \Delta s_j^{\phi,\theta} \sin \theta \, d\theta d\phi$ is averaged over all the possible orientations of the bond in the several crystallites with respect to the magnetic field and the x-ray polarization vector.

A combined analysis of the EXAFS and DiffEXAFS data has been carried out for all samples. The analysis procedure was as follows. Firstly, the experimental EXAFS spectrum is conventionally fitted using the FEFF8.10 programme [29] to determine which scattering paths in the sample are significant and to generate their amplitude $A_j(k)$ and phase $\varphi_j(k)$ information. Once a good conventional EXAFS fit was obtained, the only remaining parameters for the DiffEXAFS analysis are the structural strains $\Delta s_j'$ that were fitted to the experimental DiffEXAFS spectra. We have also performed an empirical analysis of the DiffEXAFS data using experimental backscattering $A_j(k)$ and $\varphi_j(k)$ for the isolated first and second coordination single-shells retrieved by Fourier filtering its related EXAFS spectra. Both methods of analysis gave consistent results so we only present the results from the first more accurate analysis.

Since the dependence of the change in $s_j$ with the magnetic field is not known, we can either directly compare the local magnetostriction strains determined by the DiffEXAFS analysis with the macroscopic magnetostriction values measured in the polycrystalline samples or, following the approach used by J. Díaz et al. in Ref. [18], we can associate a specific $j^{th}$ coordination shell in EXAFS with a specific crystal direction. Then, we can define a magnetostriction coefficient $\lambda_{Dj}$ measured by DiffEXAFS as

$$\lambda_{Dj} = \frac{\Delta s_j'}{s_j} \qquad (5)$$

For instance, in the case of CoFe$_2$O$_4$ and La$_{0.5}$Sr$_{0.5}$CoO$_3$ samples, the first oxygen coordination shell around the octahedral Co atoms is related to the (100) direction so the magnetostriction strain of the Co-O bond from the DiffEXAFS analysis will allow us to extract the $\lambda_{D100}$ to be compared to the macroscopic $\lambda_{100}$ one.

In the comparison with the macroscopic magnetostriction coefficient $\lambda_j$ along a crystal direction, it is important to remember that, for a polycrystal, DiffEXAFS gives the magnetostriction coefficient $\lambda_{Dj}$ of a specific crystal direction but averaged over all the possible orientations of the crystal with respect to the magnetic field direction and the polarization of the beam (see Eqs. (4) and (5)). Díaz et al. [18] have calculated the relationship between the strain $\Delta s'_j$ measured by DiffEXAFS for the (100) and (111) directions of a cubic polycrystal (Eq. (5)) and the strain measured in a cubic single crystal when it changes from the demagnetized state to saturation by applying a magnetic field parallel to the (100) or (111) directions, respectively. For that, it is assumed that $\lambda_{Dj}$ can be calculated by the same expression used for the macroscopic magnetostriction coefficient $\lambda_j$ [30]:

$$\lambda_j = \frac{3}{2}\lambda_{100}\left(\alpha_1^2\beta_1^2 + \alpha_2^2\beta_2^2 + \alpha_3^2\beta_3^2 - \frac{1}{3}\right) + 3\lambda_{111}(\alpha_1\alpha_2\beta_1\beta_2 + \alpha_2\alpha_3\beta_2\beta_3 + \alpha_1\alpha_3\beta_1\beta_3)$$

where $\alpha_i$ and $\beta_i$ are the cosines that define the magnetic field and the measurement direction, respectively, relative to the cubic crystal axes in spherical coordinates. They found $\lambda_{D111} = \frac{2}{3} \cdot \lambda_{111}$ and $\lambda_{D100} = 0.675 \cdot \lambda_{100}$. By this approach we have compared our local magnetostriction with the macroscopic one.

### A. CoFe$_2$O$_4$

The Co K-edge EXAFS spectrum at 40 K has been fitted assuming that Co atom is located at the octahedral site of the spinel structure as it is well known [31]. The fit that better adjusted the data in the k range available for the DiffEXAFS spectra, from 2 to 10 Å$^{-1}$, included single scattering paths up to the fourth coordination shell assuming a cubic *Fd-3m* lattice. The parameters obtained from the best fit are shown in Table I and the comparison between the EXAFS spectrum at the Co K-edge and this best fit is reported in the supplementary information (Fig. S1).

**Table I**. Structural parameters of the dominant contributing coordination shells used for the fit of the EXAFS spectrum at the Co K-edge of the CoFe$_2$O$_4$ sample. The shells are numbered in increasing distance from the absorber. $s_j = 2R_j$ is the scattering path length between the absorbing Co atom and an atom in the j coordination shell and $\sigma_j^2$ is its Debye-Waller factor.

$\Delta s'_j$ is the average change in the j scattering path length for the two orientations of the applied magnetic field used for the fit of the DiffEXAFS spectrum. * Relative to the Co $O_h$ atom at the origin of the cubic *Fd-3m* cell. The number in brackets indicates the error in the last significant decimal.

| Shell j | $s_j$ (Å) | $\Delta s'_j$ (Å) | $\sigma_j^2$ (Å$^2$) | Number of legs | Scattering Path* |
|---|---|---|---|---|---|
| 1 | 4.132(26) | -0.008 | 0.009(1) | 2 | O atom at (1/4, 0, 0) |
| 2 | 5.920(22) | -0.0015 | 0.0045(7) | 2 | Co/Fe $O_h$ atom at (1/4, 1/4, 0) |
| 3 | 7.018(30) | -0.004 | 0.004(1) | 2 | Fe $T_d$ atom at (-3/8, 1/8, 1/8) |
| 4 | 10.264(32) | -0.004 | 0.0025(11) | 2 | Co/Fe $O_h$ atom at (-1/4, 1/2, 1/4) |

The Fourier transform (FT) of the DiffEXAFS signal at the Co K edge, weighted in $k^2$, for the CoFe$_2$O$_4$ sample at 40 K show almost the same features as its related EXAFS spectrum, both compared in Fig. 4. The most important difference between them is that the peaks related to the scattering of further coordination shells in the FT of the DiffEXAFS spectrum are much less intense than the first coordination shell peak. According to Eq. (4), this results in a larger strain $\Delta s_j$ for the first oxygen coordination shell around the octahedral Co atom, that is along the (100) direction. Then, the oxygen environment of Co detected by DiffEXAFS seems to be tetragonally distorted. The position of the peaks seems very similar in both FT spectra so we used the same fourth shells as in the EXAFS analysis (see Table I) to fit the DiffEXAFS data.

The best fit for the Co K-edge DiffEXAFS signal is shown in Fig. 5 and the individual contributions of the fitted strains for the different coordination shells included in the fit are shown in Fig. S2(b) (see the supplementary information). The signal from the oxygen first shell is clearly larger in amplitude than those of the second shells, resulting in a strain value $\Delta R_{CoO}/R_{CoO} \approx -1900 \times 10^{-6}$. The magnetostrictive strain for the Co-Fe$_{Td}$ scattering path is $\Delta R_{CoFe_{Td}}/R_{CoFe_{Td}} \approx -600 \times 10^{-6}$, larger than that of the Co-Co(Fe)$_{Oh}$ one, $\Delta R_{CoCo(Fe)_{Oh}}/R_{CoCo(Fe)_{Oh}} \approx -250 \times 10^{-6}$. This result agrees with the fact that the magnetostrictive second-shell peak of the DiffEXAFS $|\chi(R)|$ spectrum (shown in Fig. 4(a)) corresponds to the longer interatomic distance detected by EXAFS, that is the Co-Fe$_{Td}$ one (Table I). Moreover, the sign of all the magnetostrictive strains is negative in agreement with that of the macroscopic measurements [21-22].

A multisite-multishell analysis has been performed to fit the Fe-K edge EXAFS spectrum at 40 K due to the presence of two non-equivalent Fe atoms at the tetrahedral

(Td) and octahedral (Oh) sites of the spinel structure. The fit that better adjusted the data in the k range available for the DiffEXAFS spectra, from 2 to 10 Å$^{-1}$, included two single scattering paths (Fe$_{Td}$-O and Fe$_{Oh}$-O) for the first coordination shell considering that one-half of the Fe atoms in CoFe$_2$O$_4$ occupy the octahedral site and the other half the tetrahedral site and two single scattering paths (Fe$_{Oh}$-Fe$_{Td}$ and Fe$_{Oh}$-(Fe/Co)$_{Oh}$) for the second and third coordination shells. The contribution from the Fe$_{Td}$-Fe$_{Td}$ scattering path was found to be negligible. The parameters obtained from the best fit are shown in Table II and the comparison between the EXAFS spectrum at the Fe K-edge and this best fit is reported in the supplementary information (Fig. S3).

**Table II**. Structural parameters of the dominant contributing coordination shells used for the fit of the EXAFS spectrum at the Fe K-edge of the CoFe$_2$O$_4$ sample at 40 K. The shells are numbered in increasing distance from the absorber. $s_j = 2R_j$ is the scattering path length between the absorbing Co atom and an atom in the j coordination shell and $\sigma_j^2$ is its Debye-Waller factor. $\Delta s_j'$ is the average change in the j scattering path length for the two orientations of the applied magnetic field used for the fit of the DiffEXAFS spectrum. * Relative to the Fe atom at the origin of the cubic *Fd-3m* cell. The number in brackets indicates the error in the last significant decimal.

| Shell j | $s_j$ (Å) | $\Delta s_j'$ (Å) | $\sigma_j^2$ (Å$^2$) | Number of legs | Scattering Path Fe$_{Td}$* - X |
|---|---|---|---|---|---|
| 1 | 3.884(20) | -0.0025 | 0.007(1) | 2 | O atom at (1/8, 1/8, 1/8) |
| 2 | 6.924(10) | -0.003 | 0.007(1) | 2 | Fe(Co) O$_h$ atom at (-3/8, 1/8, 1/8) |
| 3 | 10.868(10) | -0.0042 | 0.007(1) | 2 | Fe(Co) O$_h$ atom at (-5/8, 1/8, 1/8) |
| Shell j | $s_j$ (Å) | $\Delta s_j'$ (Å) | $\sigma_j^2$ (Å$^2$) | Number of legs | Scattering Path Fe$_{Oh}$* - X |
| 1' | 3.916(28) | -0.0025 | 0.015(3) | 2 | O atom at (1/4, 0, 0) |
| 2' | 5.910(14) | -0.0014 | 0.006(1) | 2 | Fe(Co) O$_h$ atom at (1/4, 1/4, 0) |
| 3' | 10.262(22) | -0.003 | 0.004(1) | 2 | Fe(Co) O$_h$ atom at (-1/4, 1/2, 1/4) |

The Fourier transform (FT) of the DiffEXAFS signal at the Fe K edge, weighted in $k^2$, for the CoFe$_2$O$_4$ sample at 40 K show almost the same features as its related EXAFS spectrum, both compared in Fig. 6. In this case, the peaks related to the scattering of further coordination shells in the FT of the DiffEXAFS spectrum are not largely reduced with respect to the first coordination shell indicating a more homogeneous strain around the Fe atoms. Furthermore, the magnetostrictive peaks for the second and third shells correspond to the longer interatomic distances detected by EXAFS, that are the Fe$_{Oh}$-Fe$_{Td}$ ones (Table II), similarly to the result found at the Co K-edge.

We have fitted the DiffEXAFS spectra of $CoFe_2O_4$ at the Fe K-edge using the same contributing shells as for the related EXAFS spectrum. The best fit for the Fe K-edge DiffEXAFS is shown in Fig. 7 and the individual contributions of the fitted strains for the different coordination shells included in the fit are shown in Fig. S4(b) (see the supplementary information). The resulting magnetostrictive strains for the Fe atoms at 40 K are $\Delta R_{FeTdO}/R_{FeTdO} = \Delta R_{FeOhO}/R_{FeOhO} \approx -600\times10^{-6}$; $\Delta R_{FeTdFe(Co)Oh}/R_{FeTdFe(Co)Oh} \approx -400\times10^{-6}$ and $\Delta R_{FeOhCo(Fe)Oh}/R_{FeOhCo(Fe)Oh} \approx -200\times10^{-6}$. The sign of all the magnetostrictive strains is also negative in agreement with that of the Co atom and the macroscopic measurements [21-22]. At room temperature, the magnitude of the magnetostrictive strains for the Fe atoms is reduced by a factor around 2.2.

Another model can be used to fit the DiffEXAFS data at the Fe K-edge giving the same quality of the fit (see dashed line in Fig. 7). Based on the macroscopic magnetostriction constants measured in a single crystal of cobalt ferrite, the ratio $\lambda_{100}/\lambda_{111}$ is large, namely $\sim 10$, with $\lambda_{100}<0$ and $\lambda_{111}>0$ [32, 33]. Therefore, we can assume that only the octahedral Fe is magnetostrictive, which results in a strain for the first oxygen coordination shell $\Delta s'_j = -0.004$ Å at 40 K, i.e. $\Delta R_{FeOhO}/R_{FeOhO} \approx -1000\times10^{-6}$.

### B. $La_{0.5}Sr_{0.5}CoO_3$

The Co K-edge EXAFS spectrum at 130 K has been fitted in the k range available for the DiffEXAFS spectra, from 2 to 10 Å$^{-1}$, included single scattering paths up to the fourth coordination shell and the multiple scattering path Co-O-Co, assuming a cubic *Pm-3m* lattice. The parameters obtained from the best fit are shown in Table III and the comparison between the EXAFS spectrum at the Fe K-edge and this best fit is reported in the supplementary information (Fig. S5).

The Fourier transform (FT) of the DiffEXAFS signal at the Co K edge, weighted in $k^2$, for the $La_{0.5}Sr_{0.5}CoO_3$ sample at 25 K show almost the same features as its related EXAFS spectrum, both compared in Fig. 8. In this case, the peaks related to the scattering from second-neighbours shells in the FT of the DiffEXAFS spectrum are almost as intense as the first coordination shell. On the other hand, the position of the peaks seems similar in both FT spectra though the Co magnetostrictive second-coordination shell mostly corresponds to the longer interatomic distance determined by EXAFS that is the Co-Co scattering path (Table III).

Table III. Structural parameters of the dominant contributing coordination shells used for the fit of the EXAFS spectrum at the Co K-edge of the $La_{0.5}Sr_{0.5}CoO_3$ sample. The shells are numbered in increasing distance from the absorber. $s_j = 2R_j$ is the scattering path length between the absorbing Co atom and an atom in the j coordination shell and $\sigma_j^2$ is its Debye-Waller factor. $\Delta s'_j$ is the average change in the j scattering path length for the two orientations of the applied magnetic field used for the fit of the DiffEXAFS spectrum. * Relative to the Co atom at the origin of the cubic *Pm-3m* cell. The number in brackets indicates the error in the last significant decimal.

| Shell $j$ | $s_j$ (Å) | $\Delta s'_j$ (Å) | $\sigma_j^2$ (Å$^2$) | Number of legs | Scattering Path* |
|---|---|---|---|---|---|
| 1 | 3.84(5) | 0.00175 | 0.010(2) | 2 | O atom at (1/2, 0, 0) |
| 2 | 6.70(9) | 0.0015 | 0.005(7) | 2 | La/Sr atom at (1/2, 1/2, 1/2) |
| 3 | 7.64(6) | 0.0035 | 0.013(3) | 3, 4 | Co atom at (1, 0, 0) multiple scattering Co-O-Co |
| 4 | 10.75(10) | 0.003 | 0.009(5) | 2 | Co atom at (1, 1, 0) |

To fit the DiffEXAFS data we have used five shells, the same as in the EXAFS. The best fit is shown in Fig. 9 and the individual contributions of the fitted strains for the different coordination shells included in the fit are shown in Fig. S6(b) (see the supplementary information). The resulting magnetostrictive strains for the Co atoms at 25 K are $\Delta R_{CoO}/R_{CoO} \approx 450\times10^{-6}$, $\Delta R_{CoLa(Sr)}/R_{CoLa(Sr)} \approx 200\times10^{-6}$, $\Delta R_{CoOCo}/R_{CoOCo} \approx 460\times10^{-6}$ and $\Delta R_{CoCo2}/R_{CoCo2} \approx 300\times10^{-6}$. The sign of all the magnetostrictive strains is positive in agreement with that of the macroscopic measurements [5]. We note that the main contributions come from the O first shell and the Co second shell, in particular from the multiple-scattering path 3 in Table III. This means that the main strain occurs along the [100] direction.

V. **Discussion and Conclusions**

We have shown that the main effect induced by the magnetic field is a strain in the local environment of either the Co atom in $La_{0.5}Sr_{0.5}CoO_3$ or the Co/Fe atoms in $CoFe_2O_4$. Local magnetostriction at the Co and Fe environments in $CoFe_2O_4$ is negative i.e., the magnetostrictive lattice shrunk in the direction of the magnetic field as for the macroscopic magnetostriction [21-22]. The strains at the Co and Fe local environments are larger for the first oxygen-shell than for higher-order shells. Besides, they are clearly different for the two atomic environments. The larger strain measured by DiffEXAFS corresponds to the first oxygen coordination shell around the Co atom, being about 2-3 times larger than the one around the Fe atom at low temperatures (40 K). We also note that the strain around the Fe atoms decreases at room temperature, whereas the Co site remains almost unaltered. Therefore, the larger strain of the Co-O shell indicates that the atom responsible of the large magnetostriction in $CoFe_2O_4$ is cobalt.

Since in $CoFe_2O_4$, the oxygen atoms coordinated to Co are located at the lattice sites along the (100) crystallographic directions in the cubic crystal (Table I), we can calculate an atomic magnetostriction coefficient $\lambda_{100}^{Co}$ by applying the relationship $\lambda_{D100} = 0.675 \cdot \lambda_{100}$, where $\lambda_{D100}^{Co} = \frac{\Delta R_{Co-O}}{R_{Co-O}}$, resulting in $\lambda_{110}^{Co} \approx$ -2800 ppm. This value is markedly higher than those obtained from the macroscopic measurement in single crystals either at room temperature $\lambda_{100} \approx$ -500 ppm or at 2 K, $\lambda_{100} \approx$ -730 ppm [7]. However, the values of the DiffEXAFS strains $\Delta R/R$ for higher-order coordination shells around Co decrease with increasing the interatomic distance yielding values much more comparable to the macroscopic magnetostriction for polycrystalline $CoFe_2O_4$ that is reported to vary between -200 and -400 ppm [21, 22].

Another important point to be discussed is the role of the Fe atoms in the magnetostrictive properties of $CoFe_2O_4$. In this case, we have two different Fe sites: (1) the octahedral Fe site where the oxygen atoms are located at the lattice sites along the (1, 0, 0) crystallographic directions and (2) the tetrahedral Fe sites, where the oxygen atoms are located at the lattice sites along the (1, 1, 1) direction. Therefore, applying the same procedure as for the analysis of the Co DiffEXAFS data, we might deduce the atomic magnetostriction coefficients $\lambda_{100}^{Fe}$ and $\lambda_{111}^{Fe}$ from the measured strain around the octahedral and tetrahedral Fe sites, respectively. Our analysis was not able to differentiate between a model where both Fe sites were magnetostrictive with comparable strain and the model where only the octahedral Fe sites are considered to contribute to the magnetostriction. The latter model is better supported by the macroscopic measurements [32, 33] and results in $\lambda_{100}^{Fe} = (\frac{\Delta R_{FeOh-O}}{R_{FeOh-O}})/0.675 \approx$ -1500 ppm at 40 K and $\lambda_{100}^{Fe} = (\frac{\Delta R_{FeOh-O}}{R_{FeOh-O}})/0.675 \approx$ -600 ppm at room temperature. The magnetostriction coefficient for the Fe atoms is also higher than the macroscopic one along the (100) direction but it approaches to it more than the one for the Co atoms. Moreover, the temperature dependence reported for the macroscopic $\lambda_{100}$ coefficient [7] is due to the magnetostrictive environments of the Fe atoms.

From the above comparison we can reach several conclusions. First, locally, the strain displacement produced by the magnetic field is not uniform in the cobalt ferrite, challenging the application of the conventional theory of magnetostriction based on the crystal elastic constants for this system. This fact could be the reason for the difference in the magnitude of the atomic magnetostriction coefficients obtained by DiffEXAFS compared to the macroscopic measurements, in particular for the Co atoms. Second, the magnetostriction in $CoFe_2O_4$ is mainly conducted by the presence of high-spin $Co^{2+}$ ($3d^7 \, t_{2g}^5 e_g^2$, S=3/2) in the octahedral sites. For $Co^{2+}$, the distortion is expected to be tetragonal with $c/a<1$, characteristic of a $t_{2g}$ degeneracy, which will produce large negative magnetostrictive effects through spin-orbit-lattice interactions with an axially-distorted cubic crystal field [11]. This distortion propagates to second neighbours (included the Fe sites) but reducing their value and approaching the macroscopic coefficient [21, 22].

On the other hand, the sign of the magnetostrictive coefficient of Co, detected by DiffEXAFS, in the $La_{0.5}Sr_{0.5}CoO_3$ sample is positive, indeed opposite to that found in $CoFe_2O_4$, but in agreement with the macroscopic anisotropic magnetotostriction [3]. The detected bond strain values at 25 K are $\frac{\Delta R_{Co-O}}{R_{Co-O}} \approx 450$ ppm and $\frac{\Delta R_{Co-O-Co}}{R_{Co-O-Co}} \approx 460$ ppm so the largest magnetostriction effects is observed along the Co-O-Co bond direction, i.e. the (100) direction. Thus, the related atomic magnetostriction coefficient $\lambda_{100}^{Co}$ is $\lambda_{100}^{Co} = (\frac{\Delta R_{Co-O}}{R_{Co-O}})/0.675 \approx 700$ ppm. This value of the magnetostrictive coefficient of the oxygen environment around the Co atom in $La_{0.5}Sr_{0.5}CoO_3$ is much smaller than the one found in $CoFe_2O_4$ but in this case, Co atom is in a mixed valence state ($Co^{3+}/Co^{4+}$) and this muti-valence condition must be discussed in the interpretation of the local magnetostriction results. The positive magnetostriction indicates tetragonally distorted octahedral sites with a $c$-axis expansion ($c/a>1$) characteristic for an $e_g$ degeneracy [34, 35]. Thus, $Co^{3+}$ and $Co^{4+}$ seem to be in an intermediate-spin state, confirming the proposed model from macroscopic measurements [5]. From the Co-La/Sr bond strain, we can obtain the atomic magnetostriction coefficient along the (111) direction that is $\lambda_{111}^{Co} = (\frac{\Delta R_{Co-La(Sr)}}{R_{Co-La(Sr)}})/0.667 \approx 340$ ppm. The macroscopic coefficient for the polycrystalline sample can be then calculated from $\lambda_{111}^{Co}$ and $\lambda_{100}^{Co}$ using Eq. (7) that yields $\lambda_M$= 470 ppm.

$$\lambda_M = \frac{2\lambda_{100M}+3\lambda_{111M}}{5} \qquad (7)$$

This value is almost the same as the measured by the macroscopic measurements at 25 K and 2 T applied magnetic field [5]. This indicates that for $La_{0.5}Sr_{0.5}CoO_3$ the mechanical response to the strain produced by the magnetic field is uniform.

In summary, we have determined the atomic environments responsible for the magnetostrictive properties of polycrystalline $CoFe_2O_4$ and $La_{0.5}Sr_{0.5}CoO_3$ ferromagnetic oxides by means of the DiffEXAFS technique. The analysis of the DiffEXAFS spectra shows that the magnetostriction effect in these cobalt oxides is governed by the magnetostrictive environments around the octahedral Co atoms. The different magnetostrictive environment of Co for the two oxides is explained in terms of the different valence and spin state of the Co atom. For the $CoFe_2O_4$ spinel, $Co^{2+}$ HS sites are tetragonally distorted with a *c*-axis contraction giving rise to a large negative magnetostriction strain for the first oxygen coordination shell along the (100) direction that is independent of the temperature. This distortion is not uniform in the sample. It propagates to the neighbours Fe atoms but it is reduced in magnitude and decreases with increasing the temperature. On the other hand, a *c*-axis expanded tetragonal distortion is found for the octahedral mixed-valent intermediate spin $Co^{3+}/Co^{4+}$ atom in the $La_{0.5}Sr_{0.5}CoO_3$ perovskite, resulting in a positive magnetostriction strain along the Co-O-Co direction that is uniform in the sample in this case.

## VI. Acknowledgements

Authors would like to acknowledge the ESRF for granting beam time, besides BM23 staff for technical assistance. For financial support, we thank the Spanish Ministerio de Economía y Competitividad Project No. RTI2018-098537-C22 cofunded by ERDF from EU and Diputación General de Aragón (Project E12-17R).

**Figures**

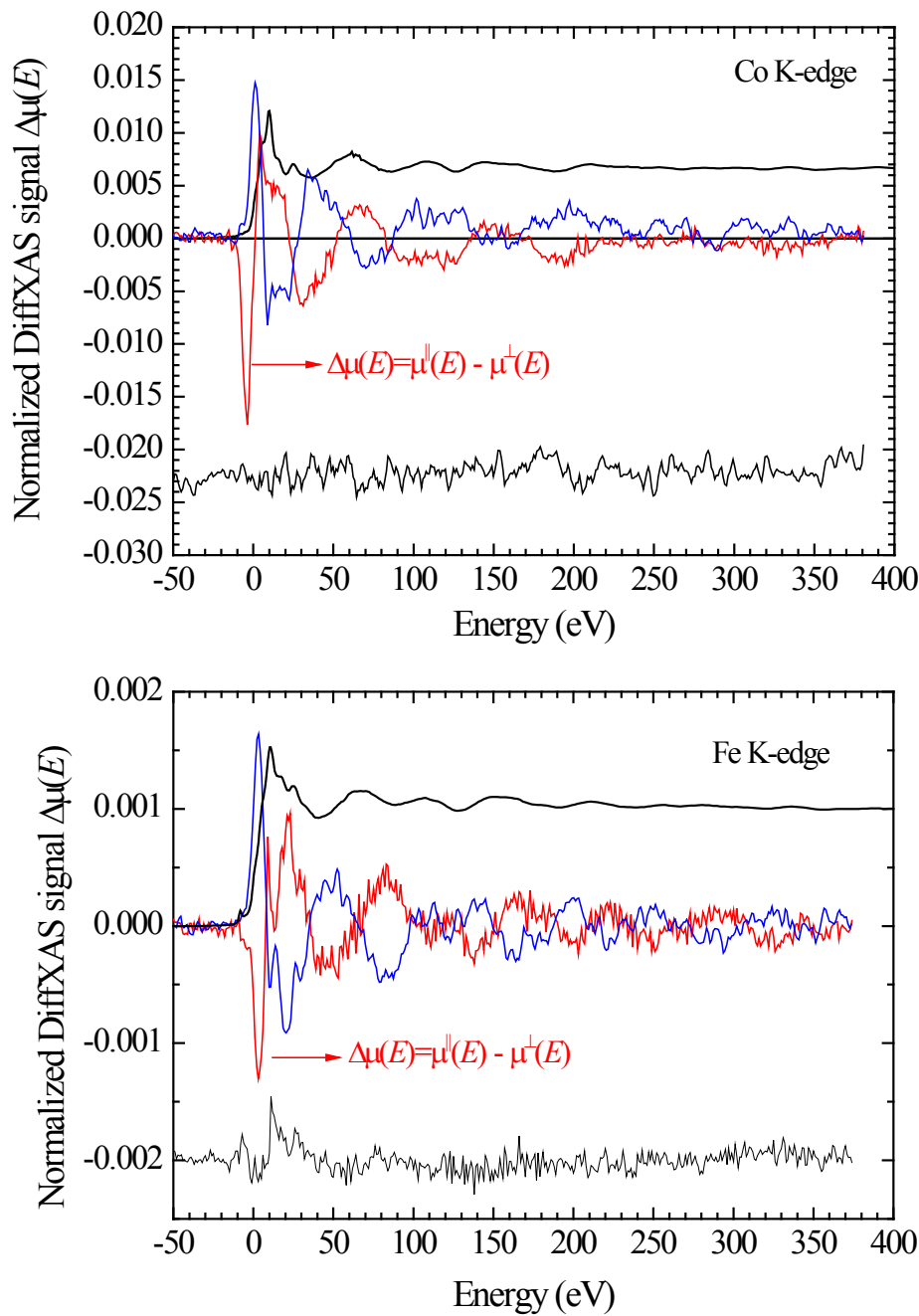

**Figure 1**. Upper panel: Normalized Co K-edge XAS $\mu(E)$ (black curve) and DiffXAS $\Delta\mu(E)$ = $\mu^{\parallel}(E) - \mu^{\perp}(E)$ (red curve) signals at room temperature for $CoFe_2O_4$. The blue curve shows the DiffXAS signal recorded with a 90º phase shift. The grey line is the DiffXAS signal measured with the magnetic field oriented at ± 45º with respect to the x-ray electric-field polarization to represent the noise. The energy axis is relative to the first inflection point of the Co K-edge (7718.5 eV). Lower panel: Normalized Fe K-edge XAS $\mu(E)$ (black curve) and DiffXAS $\Delta\mu(E)$ = $\mu^{\parallel}(E) - \mu^{\perp}(E)$ (red curve) signals at room temperature. The blue curve shows the DiffXAS signal recorded with a 90º phase shift. The grey line is the DiffXAS signal measured with the magnetic field oriented at ± 45º with respect to the x-ray electric-field polarization to represent the noise. The energy axis is relative to the first inflection point of the Fe K-edge (7122 eV).

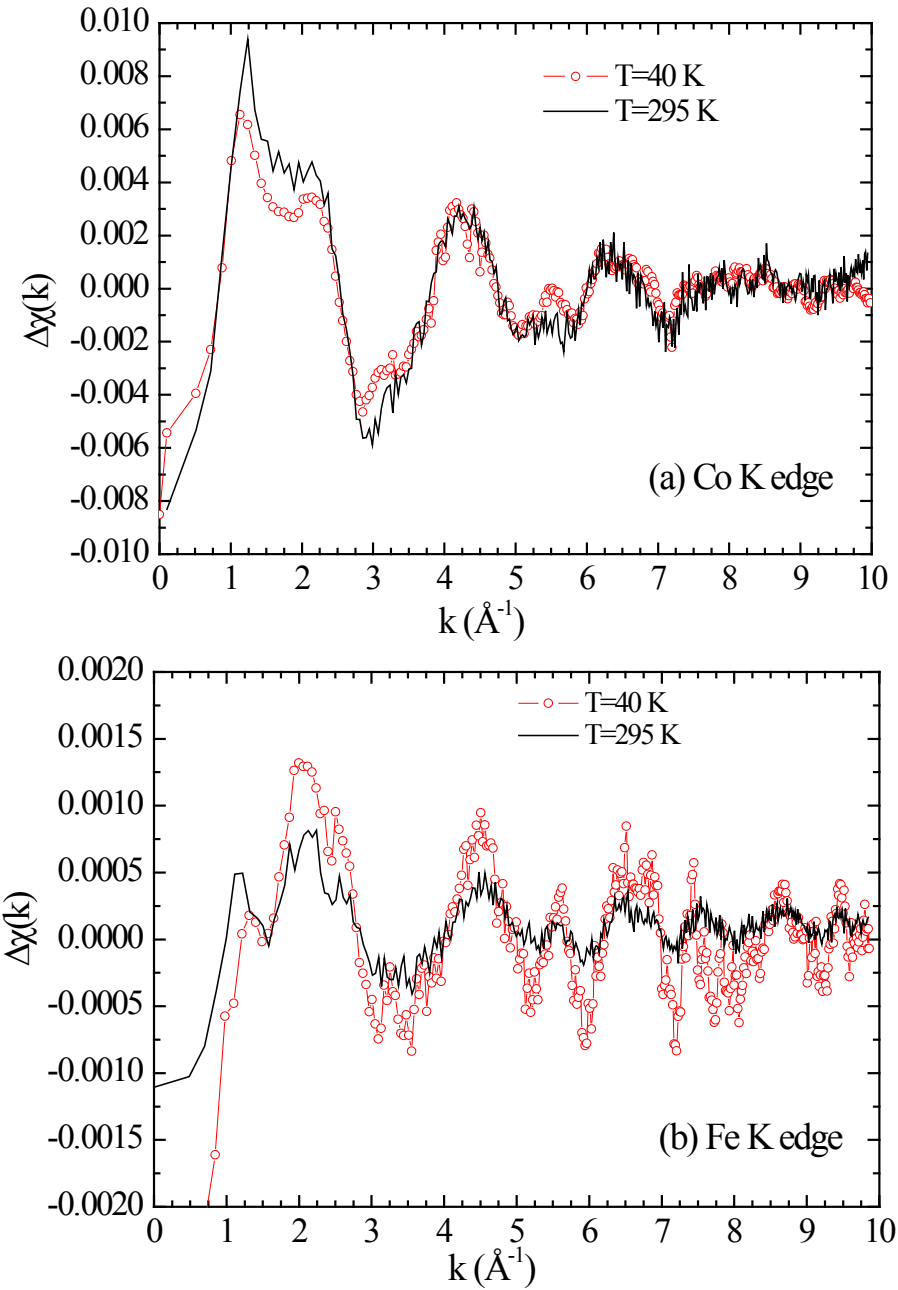

**Figure 2**. DiffEXAFS spectra of the CoFe$_2$O$_4$ sample at 40 K (red) and 295 K (black) at the Co K-edge (a) and the Fe K-edge (b).

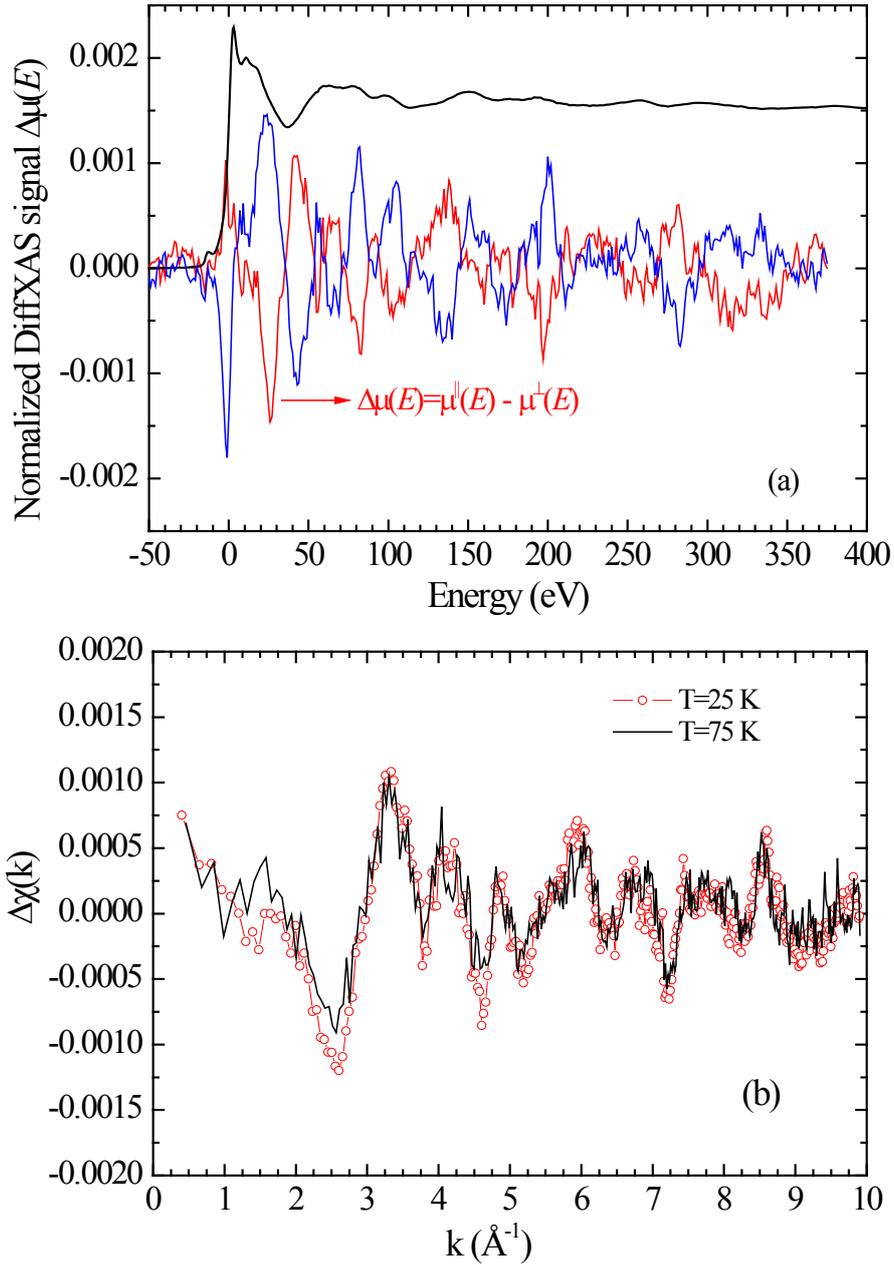

**Figure 3**. (a) Normalized Co K-edge XAS $\mu(E)$ (black curve) and DiffXAS $\Delta\mu(E) = \mu^{\parallel}(E) - \mu^{\perp}(E)$ (red curve) signals at T=25 K for the $La_{0.5}Sr_{0.5}CoO_3$ compound. The blue curve shows the DiffXAS signal recorded with a 90º phase shift. The energy axis is relative to the first inflection point of the Co K-edge (7725 eV). (b) DiffEXAFS spectra at the Co K-edge at 25 K (red line+circles) and 75 K (black solid line).

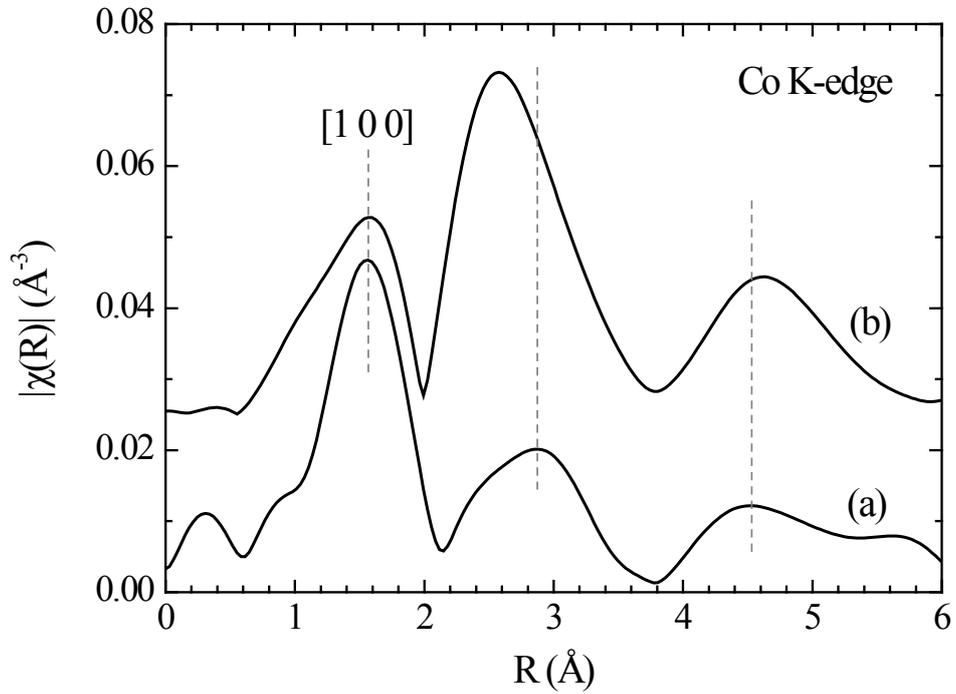

**Figure 4**. (a) Fourier transform of the DiffEXAFS signal of $CoFe_2O_4$ weighted in $k^2$ ($k$: [2, 9.5 Å$^{-1}$]) at T=40 K compared to (b) its Fourier transformed EXAFS spectrum at the Co K-edge, which was reduced in amplitude and shifted in the y-axis for comparison. The brackets at the first peak indicate the dominant strain direction.

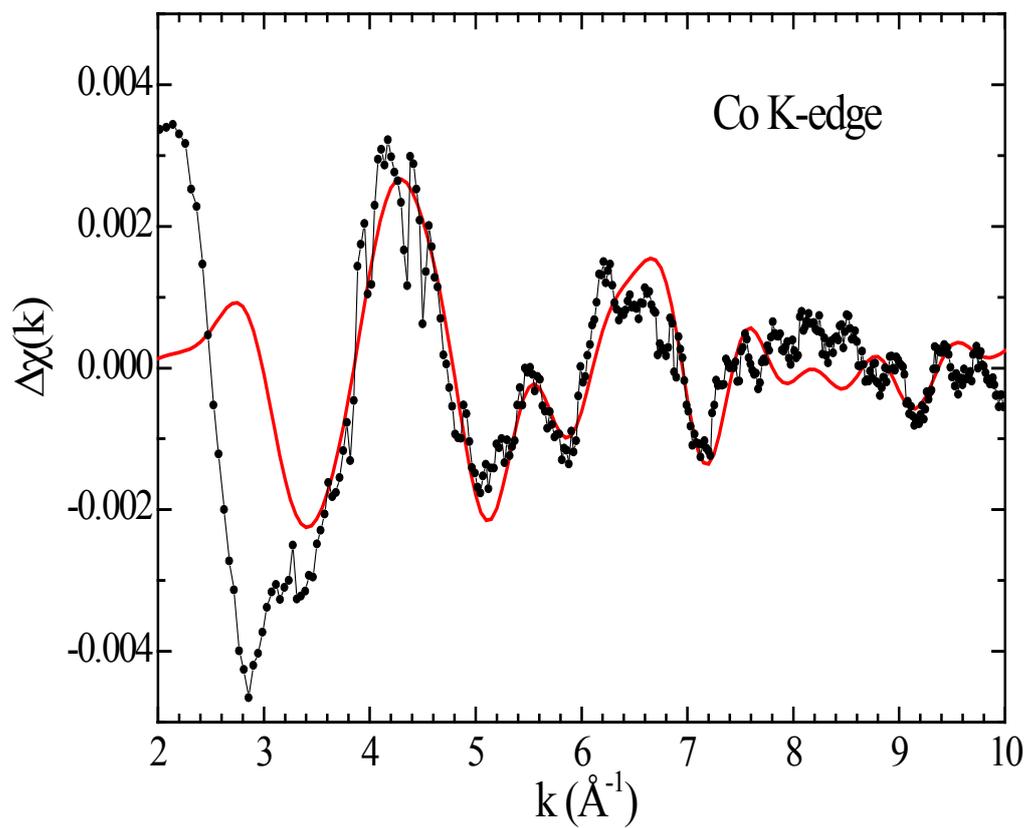

**Figure 5.** DiffEXAFS spectrum obtained at the Co K-edge of the $CoFe_2O_4$ sample at 40 K. The fit of the spectrum (red line) was done in between the $k$ limits $k$=2 and 10 Å$^{-1}$ using Eq. (4) and the parameters in Table I.

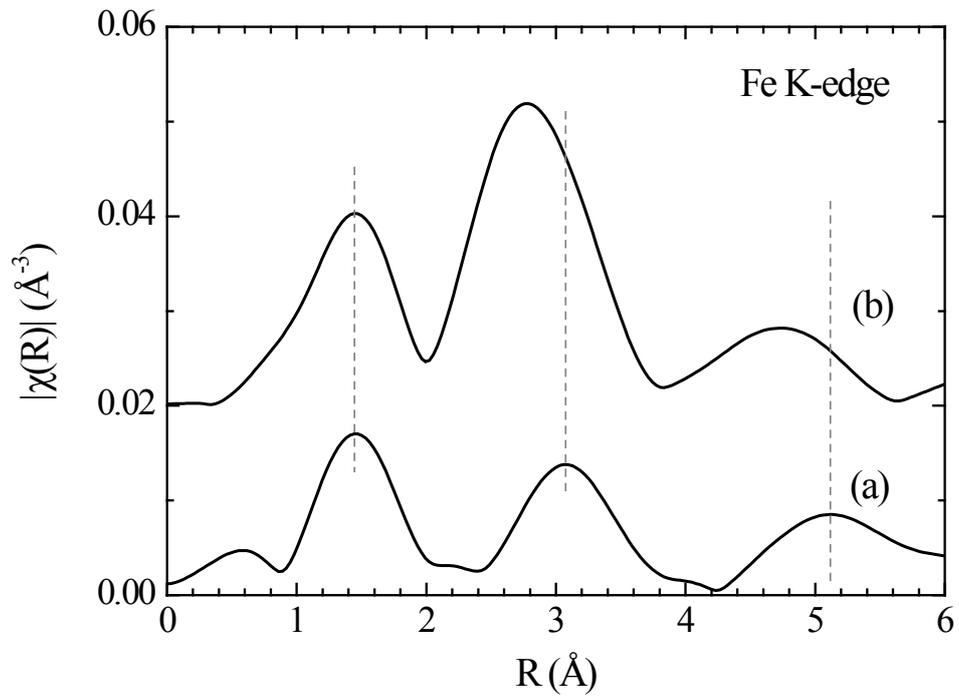

**Figure 6**. (a) Fourier transform of the DiffEXAFS signal of $CoFe_2O_4$ weighted in $k^2$ ($k$: [2, 9.5 Å$^{-1}$]) at T=40 K compared to (b) its Fourier transformed EXAFS spectrum at the Fe K-edge, which was reduced in amplitude and shifted in the y-axis for comparison.

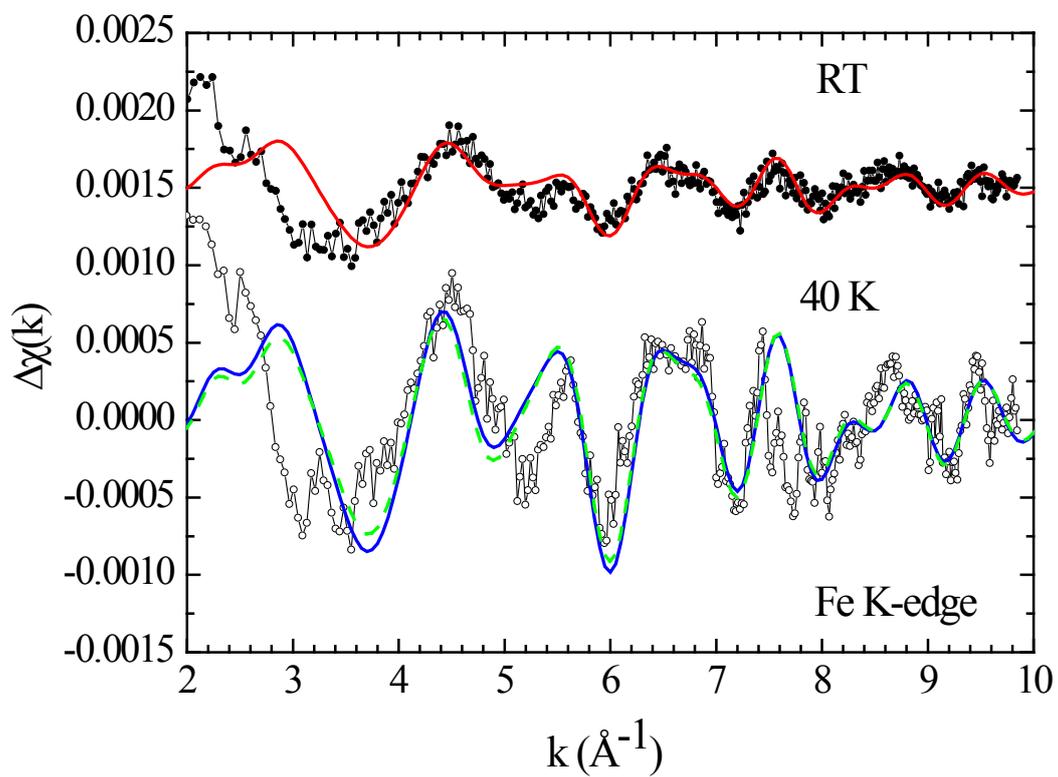

**Figure 7**. DiffEXAFS spectra at the Fe K-edge for $CoFe_2O_4$. The fit of the spectra (red line – RT and blue and grey lines – 40 K) was done in between the $k$ limits $k$=2 and 10 Å$^{-1}$ using Eq. [4]. The associated fit parameters are shown in Table II. The green dashed line is the best fit considering only the octahedral Fe contribution (1') for the first oxygen coordination shell.

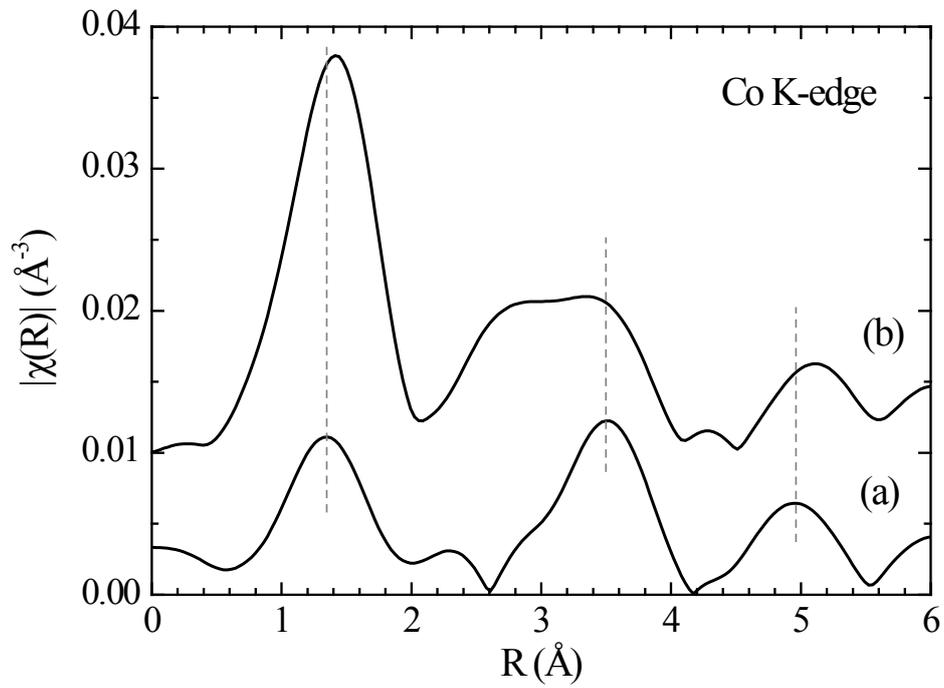

**Figure 8**. (a) Fourier transform of the DiffEXAFS signal of $La_{0.5}Sr_{0.5}CoO_3$ weighted in $k^2$ ($k$: [2, 9.5 Å$^{-1}$]) at T=25 K compared to (b) its Fourier transformed EXAFS spectrum at the Co K-edge, which was reduced in amplitude and shifted in the y-axis for comparison.

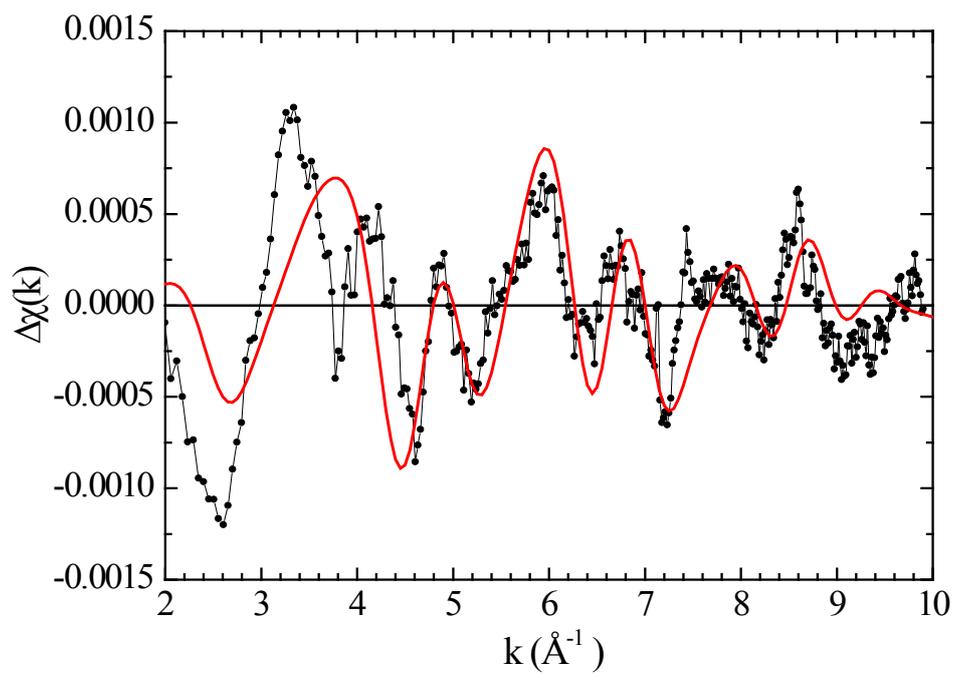

**Figure 9**. DiffEXAFS spectra at the Co K-edge for $La_{0.5}Sr_{0.5}CoO_3$ at 25 K. The fit of the spectra (red line) was done in between the $k$ limits $k$=2 and 10 Å$^{-1}$ using Eq. (4). The associated fit parameters are shown in Table III.

**Supplementary Information**

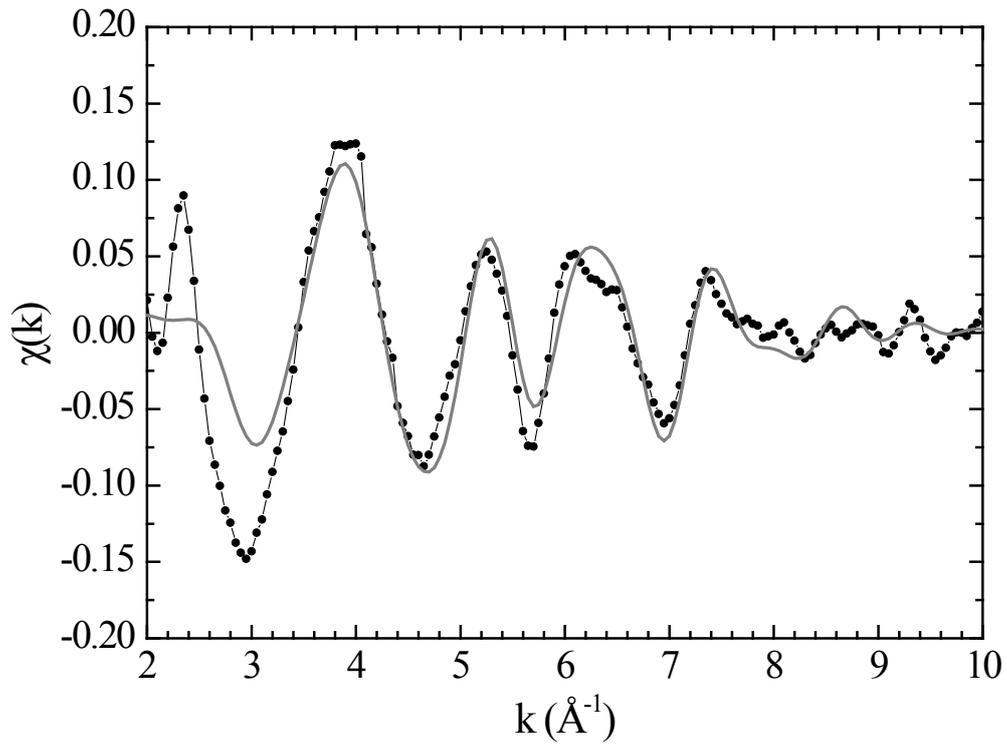

**Figure S1**. EXAFS spectrum obtained at the Co K-edge of the $CoFe_2O_4$ sample at 40 K. The grey line is the best fit to the spectrum using the parameters of Table I.

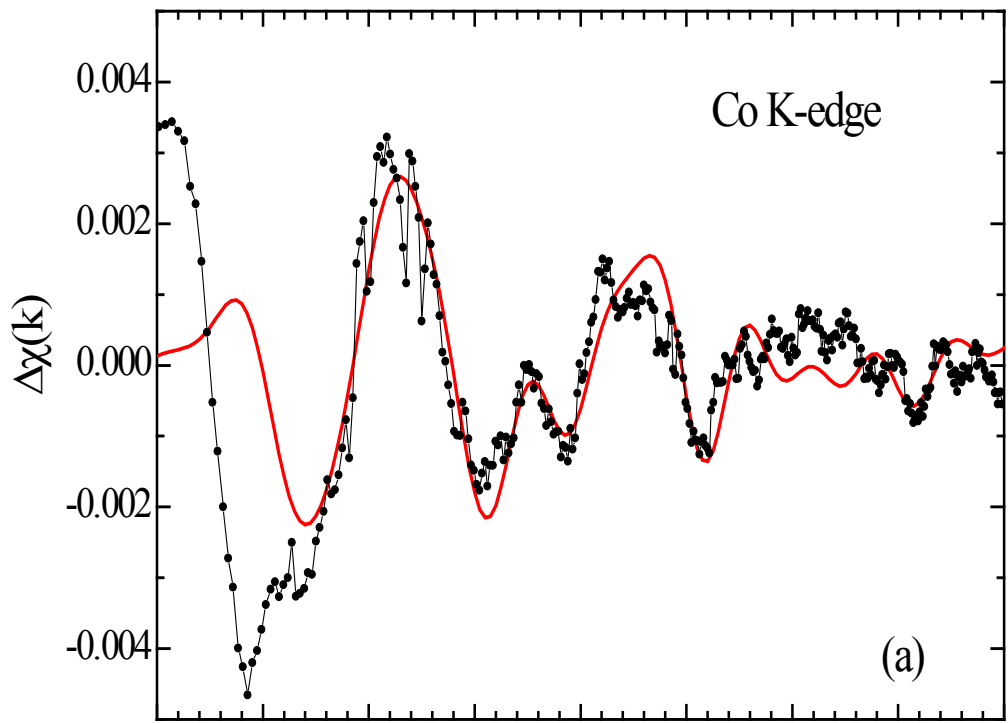

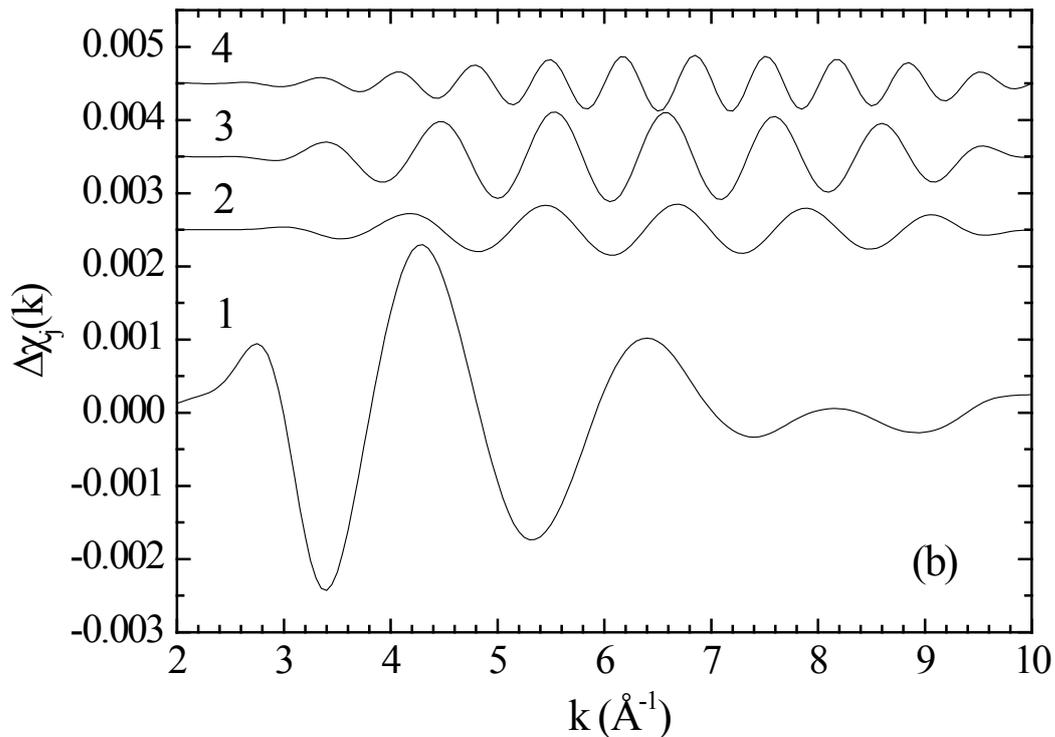

**Figure S2. (a)** DiffEXAFS spectrum obtained at the Co K-edge of the $CoFe_2O_4$ sample at 40 K. The fit of the spectrum (red line) was done in between the $k$ limits $k$=2 and 10 Å$^{-1}$ using Eq. (4) and the parameters in Table I. **(b)** The main individual contributions to the DiffEXAFS signal, with the numbers referring to the same shells as detailed in Table I. The curves have been displaced vertically from each other for clarity.

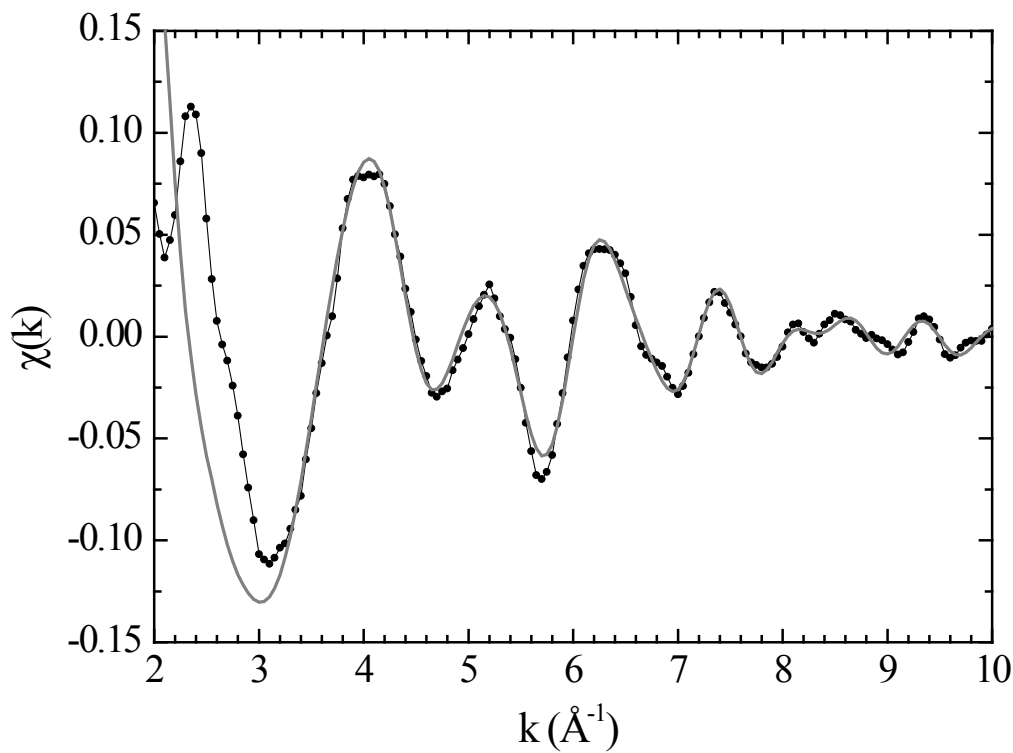

**Figure S3**. EXAFS spectrum obtained at the Fe K-edge of the $CoFe_2O_4$ sample at 40 K. The grey line is the fit to the spectrum using the parameters of Table II.

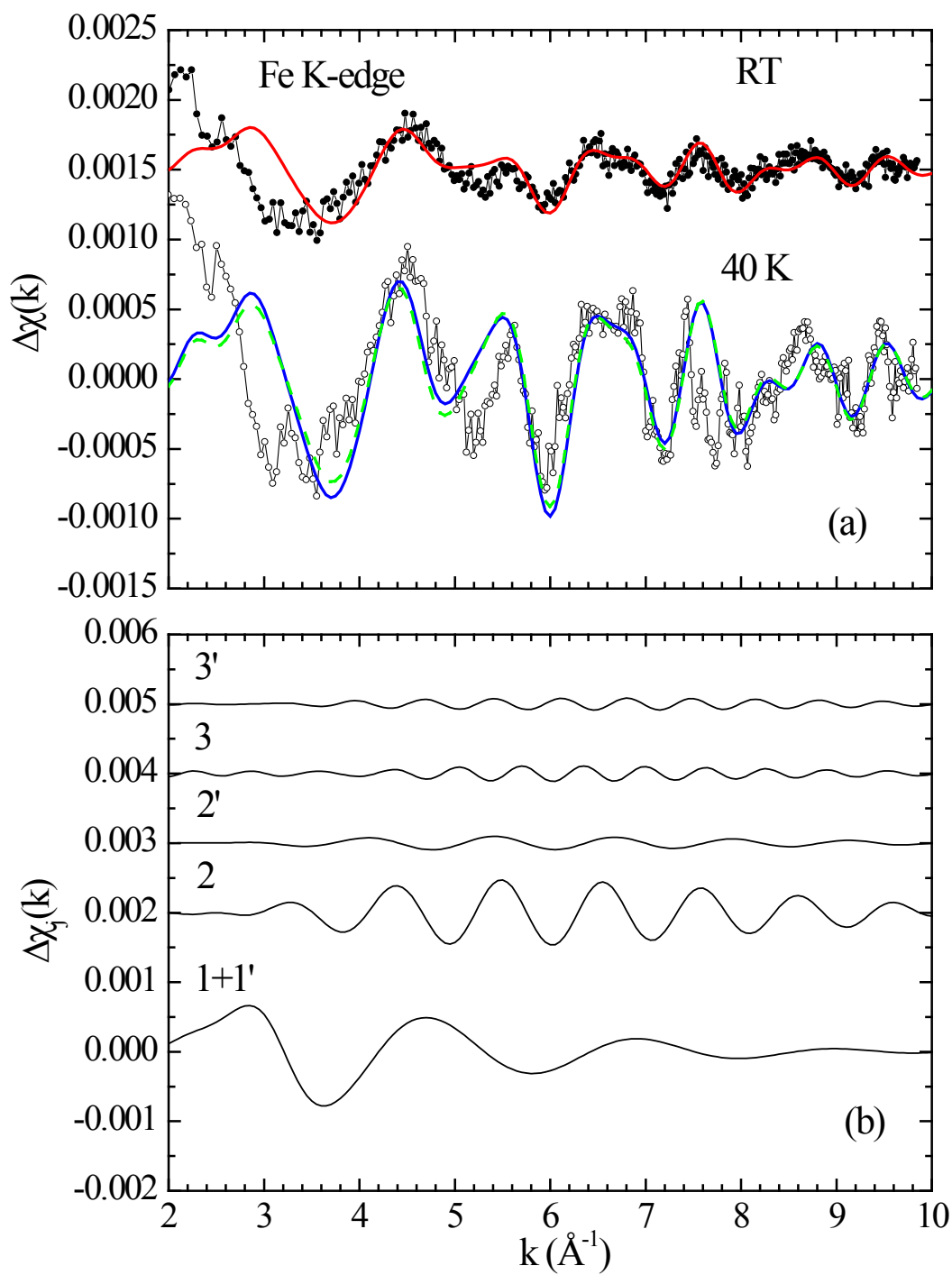

**Figure S4.** **(a)** DiffEXAFS spectra at the Fe K-edge for $CoFe_2O_4$. The fit of the spectra (red line – RT and blue and grey lines – 40 K) was done in between the $k$ limits $k=2$ and 10 Å$^{-1}$ using Eq. (4). The associated fit parameters are shown in Table II. The green dashed line is the best fit considering only the octahedral Fe contribution (1') for the first oxygen coordination shell. **(b)** The main individual contributions to the DiffEXAFS signal at 40 K, with the numbers referring to the same shells as detailed in Table II. The curves have been displaced vertically from each other for clarity.

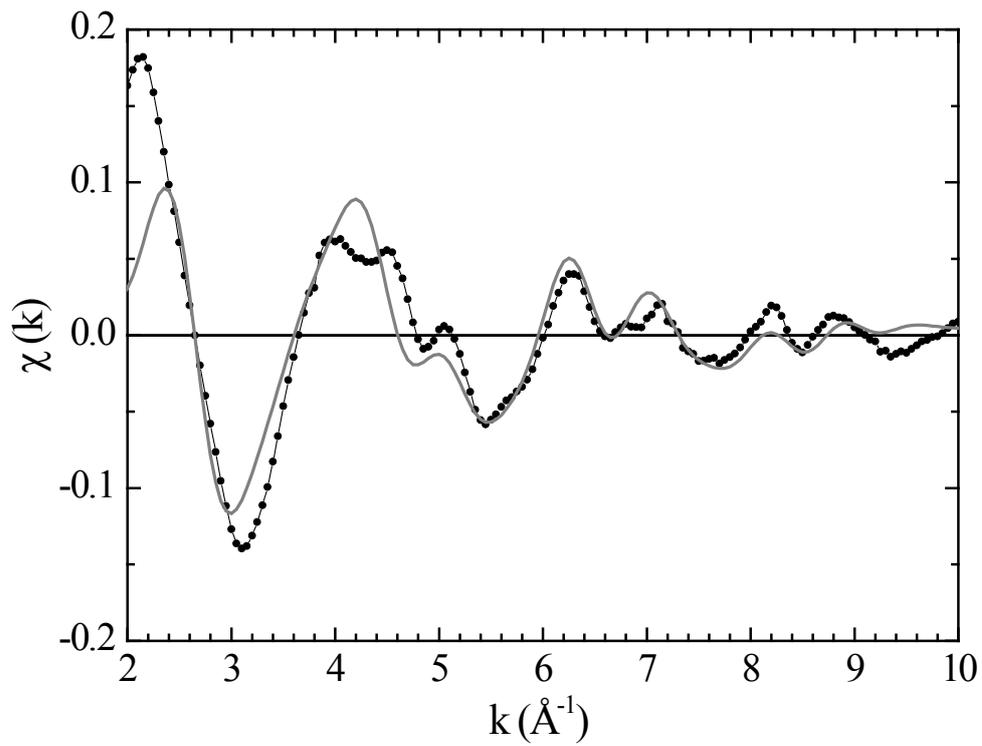

**Figure S5**. EXAFS spectrum obtained at the Co K-edge of the $La_{0.5}Sr_{0.5}CoO_3$ sample at 130 K. The grey line is the fit to the spectrum using the parameters of Table III.

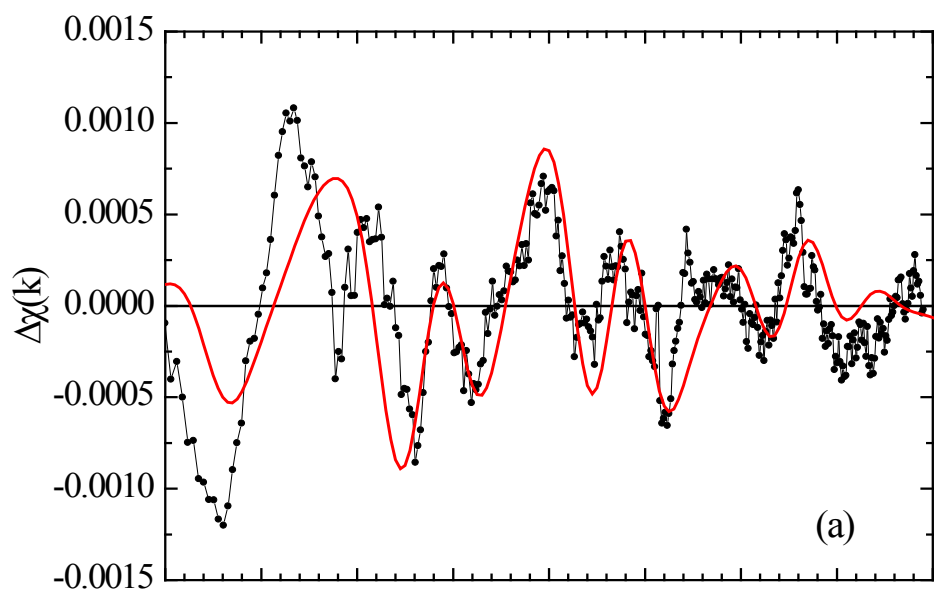

(a)

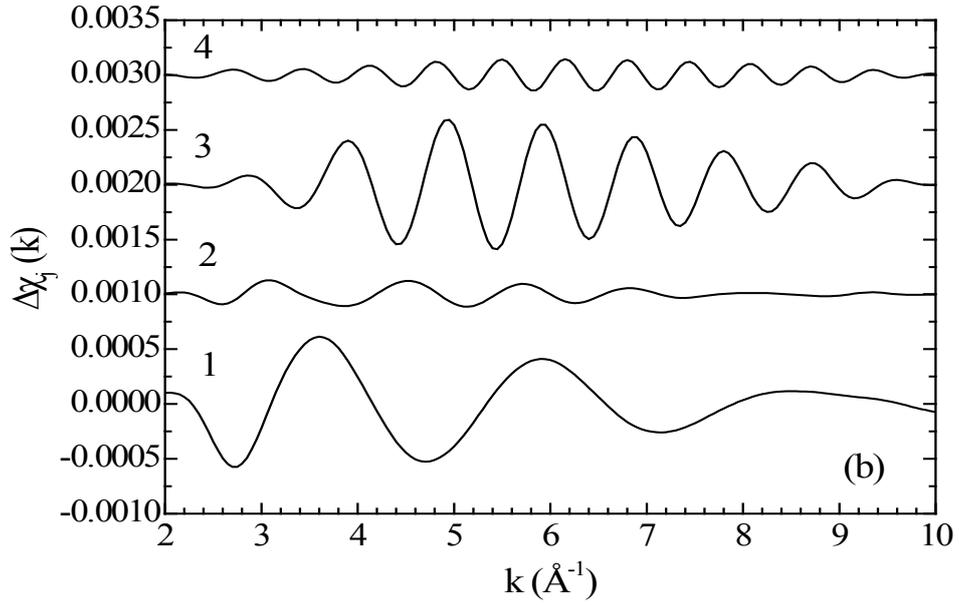

**Figure S6.** **(a)** DiffEXAFS spectra at the Co K-edge for $La_{0.5}Sr_{0.5}CoO_3$ at 25 K. The fit of the spectra (red line) was done in between the $k$ limits $k=2$ and 10 Å$^{-1}$ using Eq. (4). The associated fit parameters are shown in Table III. **(b)** The main individual contributions to the DiffEXAFS signal at 25 K, with the numbers referring to the same shells as detailed in Table III. The curves have been displaced vertically from each other for clarity.